# Demonstration of an ac Josephson junction laser

M.C. Cassidy,[1] A. Bruno,[1] S. Rubbert,[2] M. Irfan,[2] J. Kammhuber,[1] R.N. Schouten,[1,2] A.R. Akhmerov,[2] L.P. Kouwenhoven[1,2]*

[1]QuTech, Delft University of Technology, P.O. Box 5046, 2600 GA Delft, The Netherlands,
[2]Kavli Institute for Nanoscience, Delft University of Technology, P.O. Box 5046, 2600 GA Delft, The Netherlands

*To whom correspondence should be addressed; E-mail: l.p.kouwenhoven@tudelft.nl.

Superconducting electronic devices have re-emerged as contenders for both classical and quantum computing due to their fast operation speeds, low dissipation and long coherence times. An ultimate demonstration of coherence is lasing. We use one of the fundamental aspects of superconductivity, the ac Josephson effect, to demonstrate a laser made from a Josephson junction strongly coupled to a multi-mode superconducting cavity. A dc voltage bias to the junction provides a source of microwave photons, while the circuit's nonlinearity allows for efficient down-conversion of higher order Josephson frequencies down to the cavity's fundamental mode. The simple fabrication and operation allows for easy integration with a range of quantum devices, allowing for efficient on-chip generation of coherent microwave photons at low temperatures.

Josephson junctions are natural voltage to frequency converters via the AC Josephson effect. For a Josephson junction without an applied DC voltage bias, Cooper pairs tunnel coherently from one superconducting condensate to the other resulting in a supercurrent flowing without dissipation. However, for a nonzero DC voltage bias less than the superconducting energy gap, transport is prohibited unless the excess energy ($hf = 2eV_b$, where $h$ is Planks constant, $f$ is frequency and *2e* the charge on the Cooper pair) can be dissipated into the environment.

The analogy between a single Josephson junction and a two-level atom was first proposed theoretically in the 1970's (*1*). The voltage difference across the junction provides the two energy levels for the Cooper pairs, and spontaneous as well as stimulated emission and absorption were predicted to occur (*1, 2*). Emission from Josephson junctions into low-quality cavities, either constructed artificially or intrinsic to the junction's environment, in the so-called 'weak' coupling regime, has been studied extensively (*3–6*). However, due to the total output power of these systems being low, coherent radiation has not been directly demonstrated. By using a tightly confined cavity mode, coherent interaction of a single Josephson junction and the cavity can be achieved. Lasing results when the transfer rate of Cooper pairs across the junction, $\Gamma_{CP}$, exceeds the cavity lifetime, κ, of the microwave photons (Fig. 1A). Photon emission from alternative single emitters coupled to superconducting resonators has been the subject of several recent investigations (*7–9*).

We demonstrate lasing in the microwave frequency domain from a DC voltage biased Josephson junction strongly coupled to a superconducting coplanar waveguide resonator. Our device obeys several properties present in conventional optical lasers, including injection locking and frequency comb generation, with an injection locked linewidth of ≤ 1 Hz, which exceeds performance of other state of the art laser systems.

The laser consists of a halfwave coplanar waveguide (CPW) resonator with resonant frequency $f_0 \approx 5.6$ GHz made from thin (20 nm) NbTiN (Fig. 1B, C). A DC superconducting quantum interference device (SQUID), located at the electric field anti-node of the cavity, effectively acts as a single junction with Josephson energy tunable via the magnetic flux $\phi$ threading its loop: $E_J = E_{J0}|\cos(\pi\phi/\phi_0)|$, with $E_{J0} \sim 78$ GHz, and $\phi_0 = h/2e$ the superconducting flux quantum. One side of the SQUID is tied to the central conductor of the CPW, with the other end attached directly to the ground plane to enhance the coupling to the cavity. An on-chip inductor positioned at the electric field node of the cavity allows for a DC voltage bias to be applied across the SQUID (*10*). Coupling capacitors at each end of the cavity provide an input and output for microwave photons at rate $\kappa_{in}$ and $\kappa_{out}$, as in standard circuit-QED experiments (*11*). The device is mounted in a dilution refrigerator with base temperature T = 15 mK, with the magnetic flux through the SQUID tuned via a superconducting vector magnet (*12*).

We first examine the response of the device without applying any microwave power to the cavity input. At the output we measure the power spectral density, *S(f)*, of the emitted microwave radiation as a function of voltage bias $V_b$. Simultaneously, we record the corresponding current flowing through the device, $I_D = 2e\Gamma_{CP}$. The coupling between a DC voltage biased Josephson junction and a cavity, $\lambda = E_J/\phi_0^2 C\omega_0^2$, is set by the junction's Josephson energy, $E_J$, together with the cavity inductance $L = 1/C\omega_0^2$ where C is the cavity capacitance and $\omega_0 = 2\pi f_0$. When the device is configured in the weak Josephson coupling regime ($\lambda \ll 1$, Fig. 2A), by tuning the external flux close to $\phi = \phi_0$, a series of discrete microwave emission peaks are visible at bias voltages corresponding to multiples of the bare cavity resonance ($V_b = nV_r = nhf_0/2e \approx$ n × 11.62 µV). At each of these emission bursts, we observe an increase in DC current (Fig. 2B), a measure of inelastic Cooper pair transport across the Josephson junction. In this weak coupling regime, both the current and microwave emission are dominated by linear effects, with the rate of photon emission determined by the environmental impedance (*6, 13*) (Fig. 2C).

We increase the microwave emission by increasing E$_J$ via the applied flux, to the extent that the junction and cavity become strongly coupled and the system transitions to nonlinear behavior ($\lambda \gg 1$) (*14*). In contrast to the discrete emission peaks seen at low Josephson energy, the emission now shifts to higher bias voltages, persisting continuously even when the voltage bias is detuned from resonance (Fig. 2D), and is accompanied by a constant flow of Cooper pairs tunneling across the junction (Fig. 2E). The emission peaks at voltages corresponding to multiples of the cavity resonance, exhibiting bifurcations common to non-linear systems under strong driving. Between these points of instability, the emission linewidth narrows to ~ 22 kHz, well below the bare cavity linewidth of ~ 5 MHz (*12*), corresponding to a phase coherence time $\tau_c = 1/\pi\Delta f_0 = 15$ µs. The enhancement in emission originates from stimulated emission, as a larger photon number in the cavity increases the probability of reabsorption and coherent reemission by the junction. Notably, the emission power increases here by more than three orders of magnitude, while the average DC power input, $P_{in} = V_b I_D$, varies by only a factor of three. By comparing P$_{in}$ with the integrated output power, we estimate a power conversion efficiency P$_{out}$/P$_{in}$ > 0.3 (*12*), several orders of magnitude greater than achieved for single junctions without coupling to a cavity (*3*), and comparable only to arrays containing several hundred synchronized junctions (*4*). Similar power conversion efficiencies have been seen in other strongly coupled single emitter-cavity systems (*8, 9*). Application of a larger perpendicular magnetic field adjusts the cavity frequency, directly tuning the laser emission frequency by more than 50 MHz (*12*).

To understand the emission characteristics, we numerically simulate the time evolution of the coupled resonator-Josephson junction circuit for increasing $E_J$ (*12*). If the cavity only supports a single mode, the emission power at n = 1 grows with increasing $E_J$, however there is only weak emission for bias voltages corresponding to n > 1. Higher-order cavity modes allow for direct emission of higher frequency photons that can be down-converted to the fundamental cavity frequency via the non-linearity of the Josephson junction. Simulations for n = 20 modes show that for weak coupling ($\lambda \ll 1$), disconnected resonant peaks are visible in the response, in agreement with the experimental data (Fig. 2A). Only combining strong coupling ($\lambda \gg 1$) with the presence of many higher order modes do we find a continuous narrow emission line as observed in experiment (Fig. 2B). Simulations show that this behavior persists even if the mode spacing is inhomogeneous, as under strong driving the presence of harmonics and sub harmonics for each mode means that there always can always be down conversion to the frequency of the fundamental mode of the cavity (*12*).

To directly confirm lasing, we measure the emission statistics in the high Josephson coupling regime at $V_b$ = 192.5 µV. The emitted signal is mixed with an external local oscillator and the resulting quadrature components digitized with a fast acquisition card (Fig. 3A). A time series of the demodulated free running laser emission over a period of 100 µs (Fig. 3B) shows a clear sinusoidal behavior, never entering a sub-threshold state. This is in contrast to recently demonstrated lasers made from quantum dots (*9*) or superconducting charge qubits (*7*, *8*), which are strongly affected by charge noise. Instead the coherence of our system is disrupted by occasional phase slips (Inset Fig. 3 B). To quantify the effect of these phase slips, we plot the autocorrelation $g^{(1)}$ (Fig. 3C) and extract a phase coherence time of 14 µs, in good agreement with the value extracted from the free running linewidth.

To confirm coherence over longer time scales, we plot the in-phase and quadrature components of the down-converted signal from 5 x 10$^5$ samples on a two-dimensional histogram (Fig. 3D). The donut shape of the histogram confirms lasing, with the radius $A = \sqrt{\overline{N}} = 172$ representing the average coherent amplitude of the system, while the finite width $\sigma_l = \sqrt{(2\delta A^2 + N_{noise})/2}$ = 6.89 a result of amplitude fluctuations in the cavity emission $\delta A = 2.66$ broadened by the thermal noise from the amplifier chain, $N_{noise}$. When the device is not lasing ($V_b$ =18 µV in Fig. 2 (A)), we record a Gaussian peak of width $\sigma_{th} = \sqrt{N_{noise}/2}$ = 6.36, corresponding to thermal emission (Fig. 3E). To extract the photon number distribution at the output of the cavity, the contribution of thermal fluctuations due to the amplifier chain in Fig. 3E is subtracted from the emission data in Fig. 3D. The extracted distribution takes the form $p_n \propto \exp\left[-(n - \overline{N})^2 / 2\overline{N}(1 + 4\delta A^2)\right]$ centered at $\overline{N} \approx 29\,600$, (red curve in Fig 3F). In contrast, a perfectly coherent source would show a shot noise limited Poissonian distribution, which tends to a Gaussian distribution of the form $p_n \propto \exp\left[-(n - \overline{N})^2 / 2\overline{N}\right]$ in the limit of large $\overline{N}$ (blue curve in Fig. 3F). The residual fluctuations in the cavity amplitude are most probably due to $E_J$ fluctuations which change the instantaneous photon emission rate into the cavity.

Emission linewidth is a key figure of merit for lasers. A narrow linewidth implies high frequency stability and resolution, which is important for a range of technologies including spectroscopy, imaging and sensing application. One technique commonly used for stabilizing lasers is injection locking (*15, 16*) (Fig. 4A, B). The injection of a seed tone of frequency $f_{inj}$ into the cavity generates stimulated emission in the Josephson junction at this injected frequency, narrowing the emission spectrum. Figure 4C shows S(f) as a function of input power $P_{inj}$ for an injected signal with frequency $f_{inj}$ = 5.651 GHz, well within the emission bandwidth of the free running source. Linecuts at $P_{inj}$ = −127 dBm and $P_{inj}$ = −90 dBm (Fig. 4D). For very low input power, $P_{inj}$ < −140 dBm, the average photon occupation of the cavity is $\bar{N} < 1$, and the device remains unaffected by the input tone. Once the photon occupancy exceeds $\bar{N} \approx 1$, the injected microwave photons drive stimulated emission in the device, causing the emission linewidth to narrow with increasing power, reaching an ultimate (measurement limited) linewidth of 1 Hz (Fig. 4D, inset), more than 3 orders narrower than the free running emission peak, and approaching the Schawlow-Townes limit of ~ 15 mHz (*12*). In this regime, our device acts as a quantum limited amplifier, similar to other Josephson junction based amplifiers (*18, 19*), however no additional microwave pump tone is required to provide amplification. Fig. 4E shows the effect when the input tone is applied at a frequency $f_{inj}$ = 5.655 GHz, outside the cavity bandwidth. When $P_{inj}$ > −130 dBm, distortion side-bands appear at both positive and negative frequencies, and the free running emission peak is pulled towards the input tone, eventually being locked when $P_{inj}$ > −85 dBm. Linecuts at $P_{inj}$ = −127 dBm and $P_{inj}$ = −90 dBm (Fig. 4F). The positions and intensities of these emission sidebands are well described by the Adler theory for the synchronization of coupled oscillators (*19*), similar to what has been observed for both traditional and exotic laser systems (*15, 16*). The frequency range over which the device can be injection locked strongly depends on the injected power. Fig. 4 (G) shows S(f) as a function of $f_{inj}$ at an input power $P_{inj}$ = −90 dBm, showing an injection locking range Δf of almost 5 MHz. Here, the distortion sidebands span more than 100 MHz (*12*). Measurements of Δf as a function of $P_{inj}$ are shown in the inset of Fig. 4 (G). Adler's theory predicts that the injection locking range should fit $\Delta f = \alpha \sqrt{P_{inj}}$ (*19*), with a measured prefactor α = (3.66 ± 1.93) MHz/$\sqrt{W}$.

We can also use the device to generate a microwave frequency comb, an alternative to recently demonstrated four-wave mixing methods *(20)*. The time-frequency duality implies that a voltage modulation applied to the junction will create a comb in the frequency domain *(21)*. By configuring the device in the on-resonance injection locked regime ($P_{inj}$ = −110 dBm in Fig. 4C) and applying a small AC excitation of frequency $f_{mod}$ = 111 Hz to the DC bias (*12*), we generate a comb around the central pump tone with frequency separation 111 Hz. The total width of the comb is set by the amplitude of the modulation, as well as the input power of the injection lock.

Our results conclusively demonstrate lasing from a DC biased Josephson junction in the strong coupling regime. Analysis of the output emission statistics shows 15 μs of phase coherence, with no sub-threshold behavior. The Josephson junction laser does not suffer from charge-noise induced linewidth broadening inherent to semiconductor gain media, and so reaches an injection locked linewidth of < 1 Hz. The device shows frequency tunability of greater than 50 MHz by directly tuning the cavity frequency, and frequency tunability over more than 100 MHz through the generation of injection-locking sidebands. Additional frequency control may be achieved by using a broadband tunable resonator (*22*), and pulse control may be provided with a tunable coupler. The phase coherence is likely limited by fluctuations in $E_J$, either due to 1/f dependent flux noise from magnetic impurities (*23*), or due to defects within the Josephson junction, as well as thermal fluctuations in the biasing circuit that vary $V_b$. We anticipate that improvements to the magnetic shielding and passivating magnetic fluctuators, together with using a cryogenically generated voltage bias will further stabilize the emission. In this case the device would perform at the quantum limit, with a linewidth which is then only limited by residual fluctuations in the photon number in the cavity. The high efficiency together with possibility to engineer the electromagnetic environment and to guide the emitted microwaves on demand lends this system to a versatile source for propagating microwave radiation.

32. Acknowledgements. We thank D. J. van Woerkom, W. Uilhoorn and D. de Jong for experimental assistance, and C. Eichler, Y. M. Blanter, F. Hassler, J.E. Mooij, and Yu. V. Nazarov for helpful discussions. This work has been supported by the Netherlands Organisation for Scientific Research (NWO/OCW), as part of the Frontiers of Nanoscience program, Foundation for Fundamental Research on Matter (FOM), European Research Council (ERC), and Microsoft Corporation Station Q. M.C. C. and L.P.K. have filed a provisional patent application that relates to the Josephson junction laser.

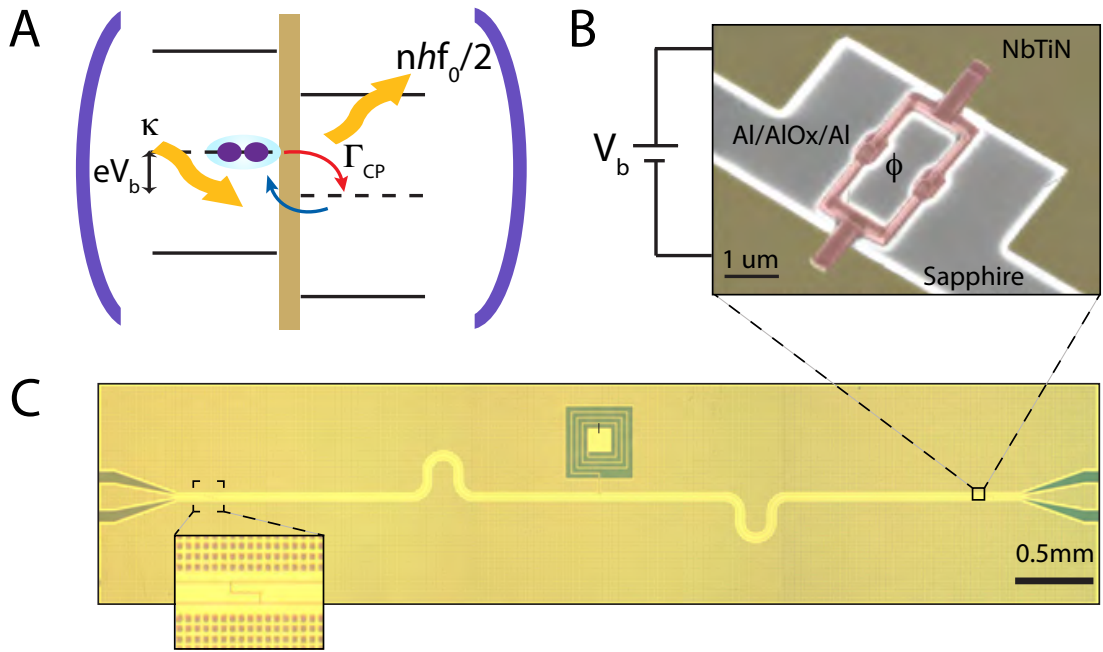

**FIGURE 1**

**Figure 1: AC Josephson laser.** **(A)** Illustration of the operating principle of the device. A DC voltage bias $V_b$ applied across the Josephson junction results in photon emission into the cavity when twice the bias voltage is equal to a multiple of the cavity frequency. If the emission rate $\Gamma_{CP}$ into the cavity exceeds the cavity lifetime $\kappa$, these cavity photons can then get reabsorbed and reemitted by the junction, a process akin to stimulated emission in atomic laser systems. Dashed lines depict the superconducting condensate, while solid lines represent the superconducting gap, $\Delta$. **(B)** Scanning electron microscope (false color) and **(C)** optical microscope image of the device. A DC SQUID (red), acting as a tunable Josephson junction, is strongly coupled to the electric field anti-node of a half-wave superconducting coplanar waveguide resonator (yellow).

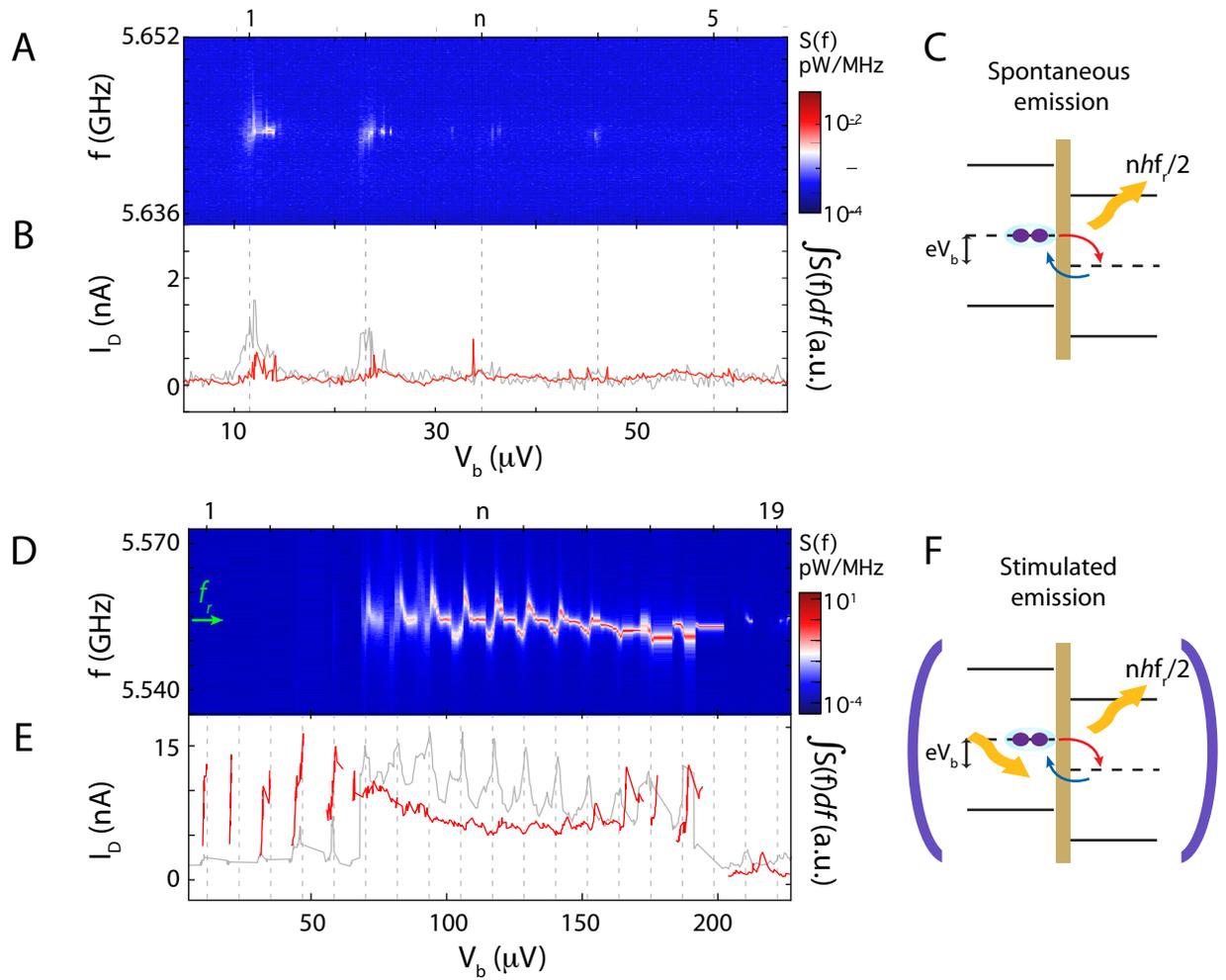

**FIGURE 2**

**Figure 2: Microwave emission from the Josephson laser. (A)** Power spectrum of the emitted radiation $S(f)$, **(B)** integrated emission (grey) and corresponding current flowing through the Josephson junction $I_D$ (red) as a function of $V_b$ when $\lambda \ll 1$. **(C)** Cooper pair transport can occur at discrete voltage biases corresponding to multiples of the cavity resonance frequency $f_0$, resulting in spontaneous photon emission into the cavity. When $\lambda > 1$ the emission **(D)** and corresponding current flow **(E)** becomes continuous across a range of bias values, peaking at bias voltages corresponding to multiples of the cavity frequency. **(F)** Cooper pair transport is accompanied by the release of multiple photons into the cavity at the fundamental frequency as well as emission of photons into the higher order resonator modes, resulting in a cavity photon occupancy large enough for stimulated emission to occur.

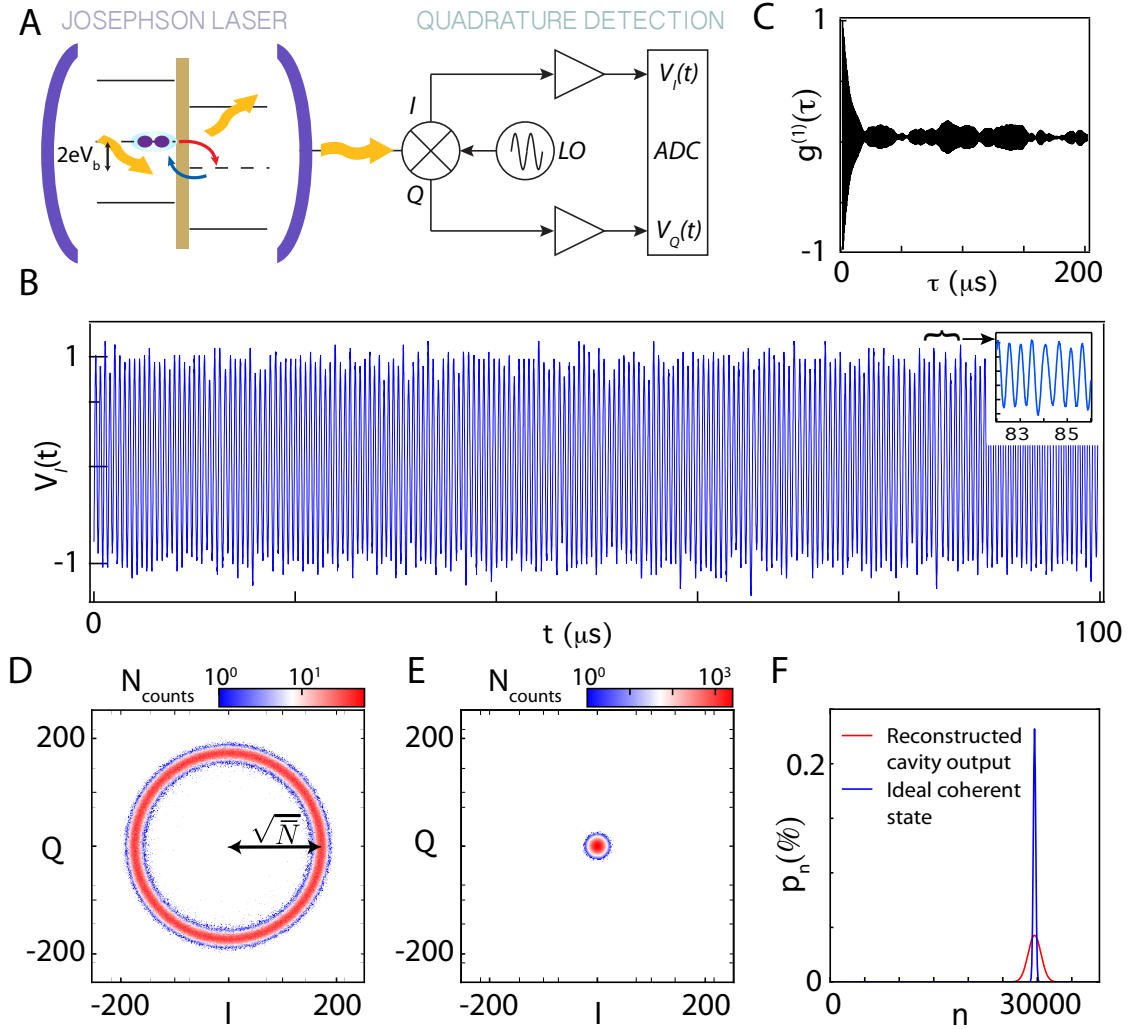

**FIGURE 3**

**Figure 3: Coherence and emission statistics of the free running Josephson laser.** (A) Real time evaluation of the emission statistics of the free running Josephson laser is performed with a heterodyne measurement setup. (B) Time series $V(t)$ of the emission over 100 μs. (Inset) Small phase slips in the emission result in a loss of coherence, resulting in an artificially broadened emission line. (C) Autocorrelation $g_{(1)}(\tau)$ of the time series shows a phase coherence of ~ 14 μs. (D) The IQ histogram acquired above threshold shows a clear donut shape, a characteristic of coherent emission. (E) The IQ histogram obtained when the device is not lasing shows a Gaussian peak centered at zero, corresponding to thermal emission. (F) The reconstructed photon number distribution at the cavity output (red) is well fit by a single Gaussian peak centered around $\bar{N} \approx 29\,600$, slightly broader than what expected for an ideal coherent source (blue).

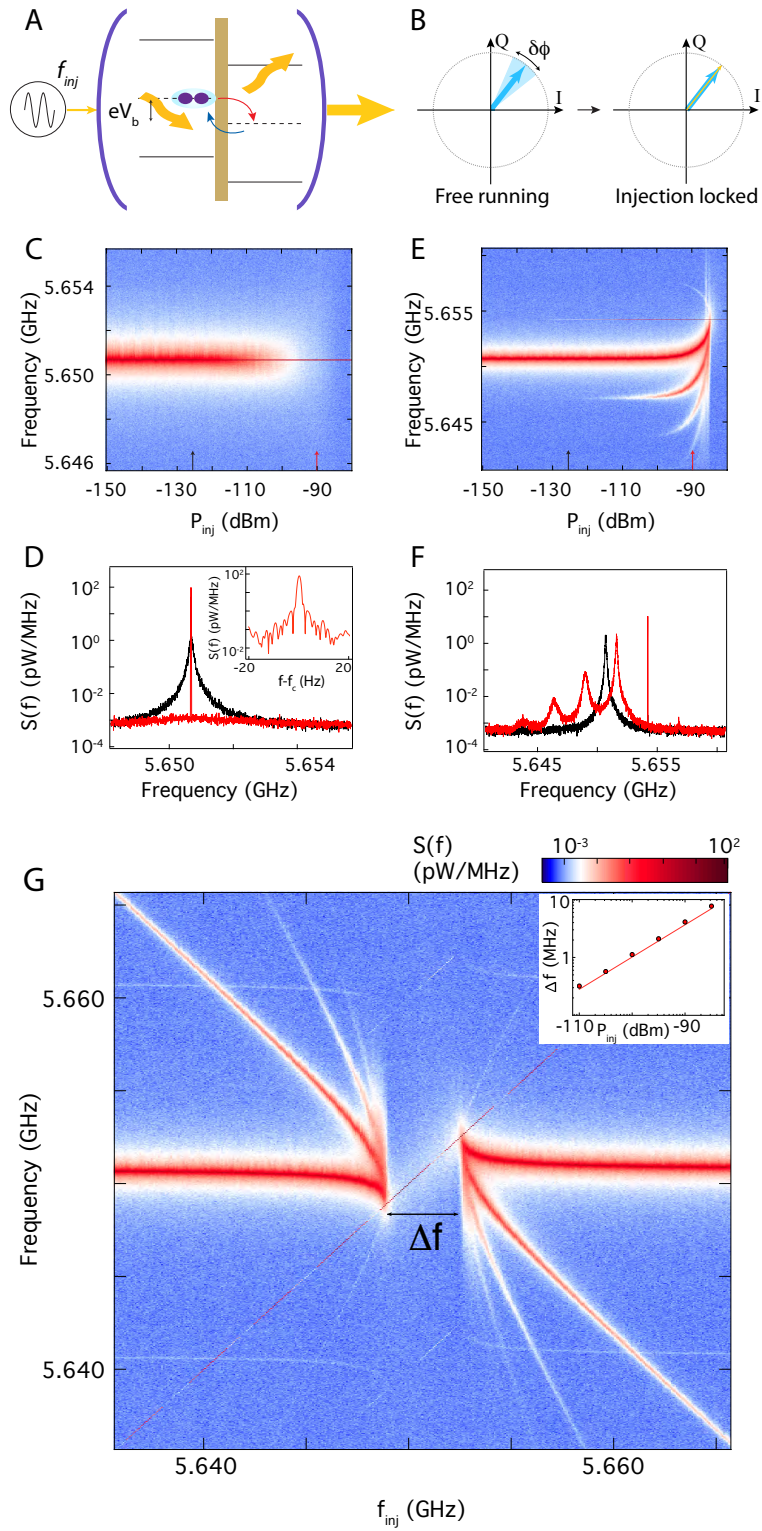

**FIGURE 4**

**Figure 4: Injection locking of the Josephson laser.** **(A)** Schematic illustration and **(B)**, phasor diagram of the injection locking process. Injection of a low power input tone into the cavity drives stimulated emission of photons synchronous with the input tone, reducing the phase fluctuations $\delta\varphi$ experienced in the free running mode. **(C)** *S(f)* as a function of $P_{inj}$ for an on-resonance input tone. **(D)** Linecuts of (C) at $P_{inj}$ = −90 dBm (red) and −127 dBm (black). (Inset) The linewidth of the injection locked laser is ≤ 1 Hz. **(E)** *S(f)* as a function of $P_{inj}$ for an off-resonance input tone, demonstrating frequency pulling. **(F)** Linecuts of (E) at $P_{inj}$ = −90 dBm (red) and −127 dBm (black). **(G)** *S(f)* at fixed input power $P_{inj}$ = −90 dBm as the frequency of the input tone $f_{inj}$ is swept. For probe frequencies $f_{inj}$ in the range Δf, the laser emission frequency locks to the frequency of the input signal. (Inset) The bandwidth of frequency locking Δf scales proportionally with the square root of the input power, in agreement with the Adler theory of coupled oscillators.